\documentclass[useAMS,usenatbib]{mn2e}
\usepackage{graphicx}
\usepackage{color}
\usepackage{amsmath}

\title[The nuclear H$_2$ and stellar discs in NGC 4258]{The molecular H$_2$ and stellar discs in the nuclear region of NGC 4258}

\author[Menezes et al.] {R.~B.~Menezes\thanks{E-mail: roberto.menezes@iag.usp.br}, Patr\'icia~da~Silva and J.~E.~Steiner\\
Instituto de Astronomia Geof\'isica e Ci\^encias Atmosf\'ericas, Universidade de S\~ao Paulo, Rua do Mat\~ao 1226, \\Cidade Universit\'aria, S\~ao Paulo, SP CEP 05508-090, Brazil}

\begin{document}

\date{Accepted 2017 September 19. Received 2017 September 19; in original form 2017 March 8}

\pagerange{\pageref{firstpage}--\pageref{lastpage}} \pubyear{2017}

\maketitle

\label{firstpage}

\begin{abstract}

NGC 4258 is an SABbc Seyfert galaxy, located at a distance of $7.2 \pm 0.3$ Mpc. This object is well known by its nearly edge-on molecular nuclear disc, located between 0.16 and 0.28 pc from the nucleus, whose H$_2$O maser emission allows a very precise measurement of the central supermassive black hole mass ($M_{\bullet}(maser) = 3.78 \pm 0.01 \times 10^7$ $M_{\sun}$). We analyse the emission line properties and the stellar kinematics in a data cube of the central region of NGC 4258, obtained, in the \textit{K} band, with the Near-Infrared Integral Field Spectrograph, at the Gemini-north telescope. The nuclear spectrum, after the starlight subtraction, shows only the H$_2 \lambda 2.1218$ $\mu$m and Br$\gamma$ emission lines, the latter having a broad component with FWHM$_{Br\gamma}(broad) = 1600 \pm 29$ km s$^{-1}$. The spatial morphology and kinematics of the H$_2 \lambda 2.1218$ $\mu$m line are consistent with a rotating molecular disc around the supermassive black hole, with an upper limit for its diameter of 15.7 pc. The inner radio jet in this object is, in projection, almost perpendicular to the H$_2$ emitting disc detected in this work, and also to the H$_2$O maser emitting disc. The main features of the maps of the stellar kinematic parameters are well reproduced by a model of a thin rotating stellar circular disc. The supermassive black hole mass provided by this dynamical modelling ($M_{\bullet}(disc) = 2.8 \pm 1.0 \times 10^7$ $M_{\sun}$) is compatible, at 1$\sigma$ level, with the precise measurement resulting from the H$_2$O maser emission.

\end{abstract}


\section{Introduction}

Supermassive black holes (SMBHs) seem to be present at the centres of all massive galaxies \citep{kor95,ric98}. Their masses show correlations with certain parameters of the host galaxy, such as the stellar velocity dispersion of the bulge (the $M-\sigma$ relation; Ferrarese \& Merritt 2000; Gebhardt et al. 2000; G\"ultekin et al. 2009). Such correlations suggest a coevolution of the SMBH and of the host galaxy, which makes the determination of the SMBH mass one of the most relevant analyses, in this context. On the other hand, it is well accepted that SMBHs increase their masses by accreting material, during  an active galactic nucleus (AGN) phase. Therefore, the study of the circumnuclear regions of AGNs is also very important, as it can reveal details of the feeding process in these objects. The use of 3D spectroscopy is particularly useful in studies involving the determination of the SMBH mass in a galaxy or the analysis of the circumnuclear region.

NGC 4258 is an SABbc galaxy, classified as Seyfert 1.9 \citep{ho97a}. \citet{kru74} analysed the ionized gas velocity field of this galaxy, using optical slit spectra, and found a position angle $PA_{galaxy} = 146\degr$ for the line of nodes and an inclination $i_{galaxy} = 72\degr$ for the disc of the galaxy. In a later study, using an H \textsc{I} velocity field, \citet{alb80} obtained the values of $PA_{galaxy} = 150\degr$ and $i_{galaxy} = 72\degr$ for these parameters. More recently, \citet{saw07} analysed higher resolution CO $J$ = 2-1 data and found $PA_{galaxy} = 160\degr$ and $i_{galaxy} = 65.6\degr$.

A peculiar morphological feature of NGC 4258 is its `anomalous arms', which start at the nucleus and seem to intercept the regular spiral arms of the galaxy. These structures were first observed in H$\alpha$ images analysed by \citet{cou61}, who noted that they are more diffuse than normal spiral arms. 1415 MHz radio images obtained by \citet{kru72} revealed that the anomalous arms are also strong radio emitters. Many other studies of these features have been made in the optical (e.g. Burbidge, Burbidge \& Prendergast 1963; Ford et al. 1986; Rubin \& Graham 1990; Cecil, Wilson \& Tully 1992; Cecil et al. 2000) and in the radio (e.g. van Albada \& van der Hulst 1982; Hummel, Krause \& Lesch 1989; Cecil et al. 2000; Hyman et al. 2001). The optical emission line ratios \citep{cec95a} of the anomalous arms, together with the X-ray spectroscopic properties of this area \citep{cec95b,wil01,yan07}, indicate the presence of shocked gas, with a temperature in the range of $10^5 - 10^7$ K. Wilson et al. (2001) proposed that the origin of the anomalous arms is related to the interaction between the jets of the AGN in NGC 4258 and the interstellar gas. The angle between the jets and the rotation axis of the galaxy is $60\degr$. The jets shock the gas, generating cocoons of X-ray emitting hot gas in the circumnuclear region. In farther areas, the jets drive motions in the low density gas of the halo, which propagate and encounter the denser gas of the galaxy's disc. As a result, a `line of damage', corresponding to the projection of the jets on the disc, is created. In this scenario, the anomalous arms correspond to this line of damage created by the jets.

The AGN in NGC 4258 has been observed in many spectral bands. Using \textit{ASCA} observations in 2-10 keV, \citet{mak94} detected a highly obscured nuclear source, with a luminosity of $L_X(2-10~keV) = 4 \times 10^{40}$ erg s$^{-1}$. Other \textit{ASCA} observations showed that the nuclear X-ray luminosity is actually variable, with time-scales of years \citep{pta99, rey00, ter02}. \citet{you04} analysed \textit{Chandra} data and concluded that the nuclear spectrum of NGC 4258 can be reproduced by a model with two components: a heavily absorbed ($N_H = 7 \times 10^{22}$ cm$^{-2}$) hard X-ray power law, of variable luminosity, and a constant thermal soft X-ray component. The photon index of the power-law component is $\Gamma = 1.4$.

\citet{wil95} analysed polarimetric data of NGC 4258 and detected broadened emission lines ($\sim 1000$ km s$^{-1}$) and a faint blue continuum, which can be reproduced by a power law with a spectral index of $\alpha = 1.1 \pm 0.2$. This is consistent with the presence of an obscured AGN in the nucleus of this galaxy.

The unresolved nuclear source corresponding to the AGN is also clearly visible in the near-infrared, as noted by \citet{cha97}, using high-resolution \textit{H}- and \textit{K}-band images obtained with the Keck telescope. The authors suggest that the detected emission is nonthermal, although the hypothesis of thermal emission, related to a possible dust torus, cannot be discarded. The infrared continuum of this nuclear source can be well reproduced by a power law, with a spectral index of 1.4 \citep{yua02}. Using high-resolution infrared images, obtained with the \textit{Hubble Space Telescope} (\textit{HST}), \citet{cha00} obtained a nuclear luminosity of $2 \times 10^8$ L$_{\sun}$, in the spectral range of 1 - 20 $\mu$m.

The compact nuclear source in NGC 4258 was observed in the radio, with the Very Large Array (VLA), by \citet{tur94}. A jet was also detected, in this spectral band, in the central region of this galaxy, extending along the north-south direction. \citet{cec00} analysed 1.46 GHz VLA images and verified that, in scales of 0.3-300 pc, the jet has a position angle $PA_{jet} = -3\degr \pm 1\degr$. In addition, the authors observed two radio hot spots, located at projected distances of 840 pc S and 1.7 kpc N from the nucleus.

\begin{figure*}
\begin{center}
  \includegraphics[scale=0.24]{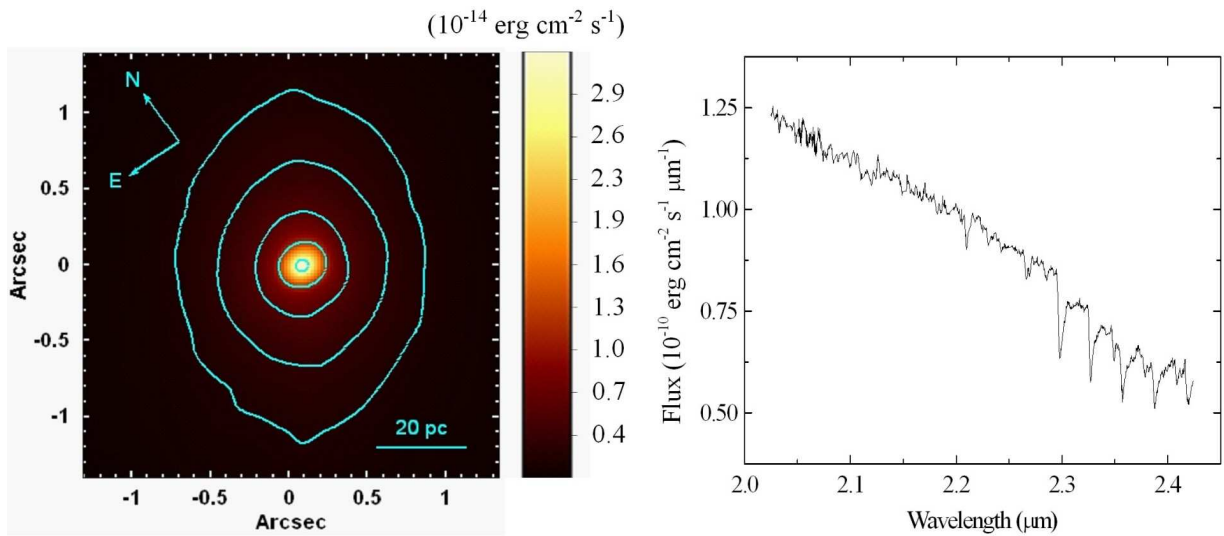}
  \caption{Left: integrated flux image of the treated data cube of NGC 4258, with the isocontours shown in cyan. Right: mean spectrum, over the entire FOV, of the treated data cube of NGC 4258.\label{fig1}}
\end{center}
\end{figure*}

NGC 4258 is famous by its H$_2$O maser emission from a geometric thin and nearly edge-on molecular nuclear disc, which is located between 0.16 and 0.28 pc from the SMBH \citep{miy95,gre95a,gre95b,her99}. The keplerian rotation curve traced by this H$_2$O maser emission allows a very precise measurement of the SMBH mass. Using observations taken with the Very Long Baseline Array, \citet{miy95} obtained an inclination of $i_{maser} = 83\degr \pm 4\degr$ for the maser emitting molecular disc and a mass of $3.6 \times 10^7$ $M_{\sun}$ for the SMBH. The position angle of this molecular disc is $PA_{maser} = 86\degr \pm 2\degr$ \citep{miy95}, so, in projection, it is almost perpendicular to the radio jet mentioned above. Based on the observed proper motions and accelerations of the maser emission, \citet{her99} calculated a geometric distance of $7.2 \pm 0.3$ Mpc to this galaxy, which also resulted in a more precise determination for the SMBH mass ($3.9 \pm 0.1 \times 10^7$ $M_{\sun}$). \citet{her05} applied a maximum likelihood analysis of the positions and velocities of the maser emission and detected a deviation from Keplerian motion in the projected rotation curve. The authors presented different models capable of explaining this discrepancy, such as the presence of a central dark cluster of objects or a massive molecular disc; however, their preferred model assumes that the maser emitting disc is inclination-warped. This scenario results in a value of $M_{\bullet}(maser) = (3.78 \pm 0.01) \times 10^7$ $M_{\sun}$ for the SMBH mass. This galaxy, together with a few others that also have nearly edge-on nuclear maser emitting discs \citep{kuo11}, has the most precise mass determinations for the central SMBH obtained so far (see Kormendy \& Ho 2013 for a review).

Other measurements of the SMBH mass in NGC 4258 have been made in later studies. \citet{sio09}, for example, using high-resolution images and spectra obtained with the \textit{HST}, together with ground-based observations, applied axisymmetric Schwarzschild models \citep{sch79} and obtained a value of $M_{\bullet} = (3.3 \pm 0.2) \times 10^7$ $M_{\sun}$ for the SMBH mass. \citet{dre15} applied anisotropic and axisymmetric Jeans models (e.g. Cappellari 2008), using a near-infrared data cube obtained with the Near-Infrared Integral Field Spectrograph (NIFS), at the Gemini-north telescope, together with \textit{HST} and 2MASS images, and obtained a mass of $M_{\bullet} = 4.8^{+0.8}_{-0.9} \times 10^7$ $M_{\sun}$ for the SMBH.  

In this paper, we analyse the emission line properties of a near-infrared data cube of the central region of NGC 4258, obtained with NIFS. We also test the validity of the model of a thin rotating stellar disc to reproduce the observed stellar kinematics of the data cube and to provide an estimate for the mass of the SMBH. Since this galaxy has one of the most precise estimates for the mass of a SMBH, it is a very suitable target to be used to perform such a methodological test. This relatively simple model of a thin rotating stellar disc may be useful in certain situations, especially because it does not require the assumption of axisymmetry. The data cube used in this work is actually the same analysed by \citet{dre15}. However, we focus our analysis on different scientific aspects. The paper is organized as follows: in Section 2, we describe the observations, the data reduction and also the data treatment. In Section 3, we analyse the emission line spectrum of the data cube. In Section 4, we present maps of the stellar kinematic parameters and try to reproduce them with the model of a thin stellar disc, in order to obtain an estimate for the mass of the SMBH. In Section 5, we discuss and compare our results with those of previous studies. Finally, we draw our conclusions in Section 6.

\section{Observations, data reduction and data treatment}

\begin{figure*}
\begin{center}
  \includegraphics[scale=0.53]{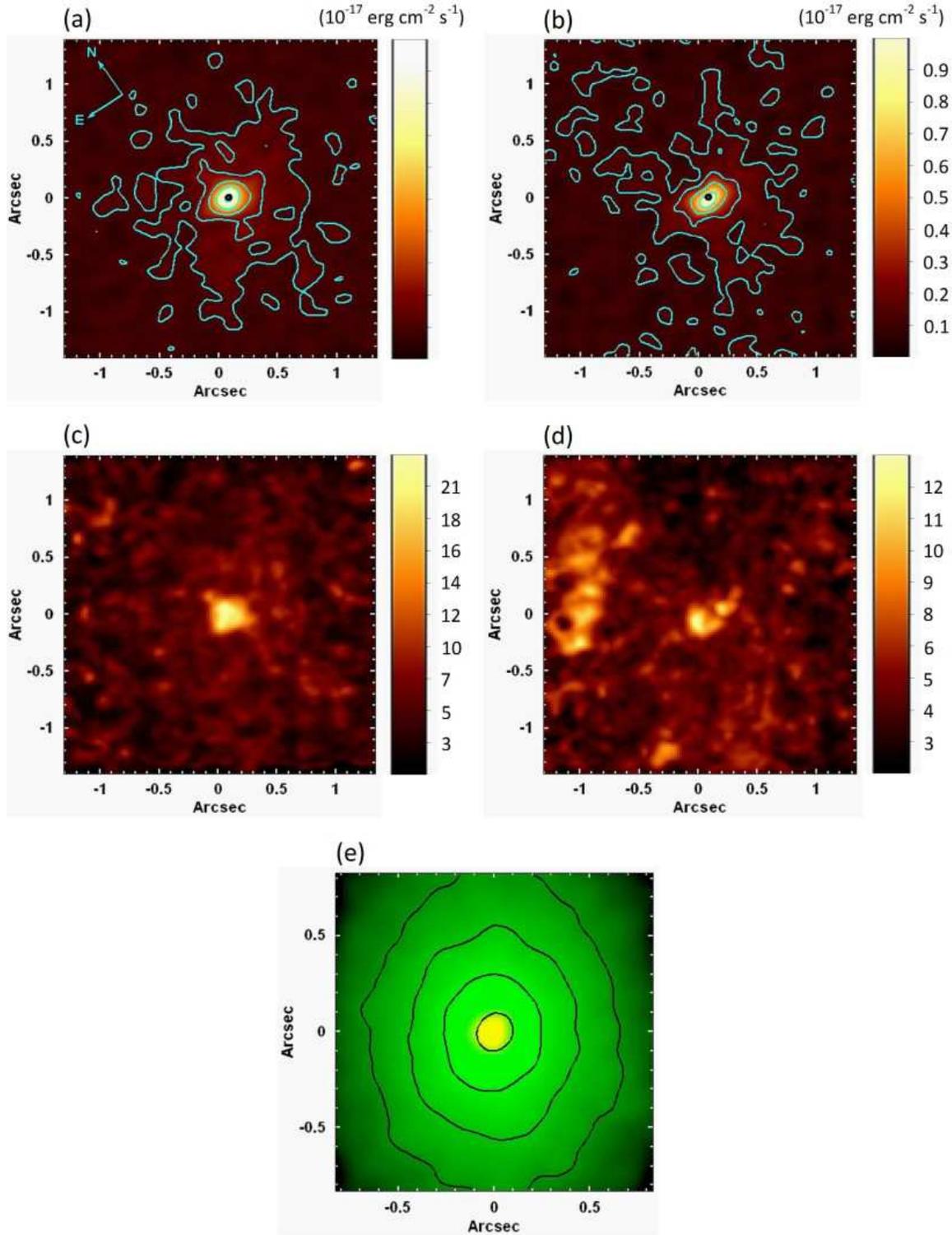}
  \caption{(a) Integrated flux image of the Br$\gamma$ emission line. (b) Integrated flux image of the H$_2 \lambda 2.1218$ $\mu$m emission line. (c) A/N ratio map of the Br$\gamma$ emission line. (d) A/N ratio map of the H$_2 \lambda 2.1218$ $\mu$m emission line. (e) RG composite, with reduced FOV, with the image of the blue wing (2.1564 - 2.1608 $\mu$m) of the broad component of Br$\gamma$ shown in red and the image of the $^{12}$CO(2-0) band (obtained by subtracting the image corresponding to the wavelength interval of the $^{12}$CO(2-0) band, 2.2929 - 2.2982 $\mu$m, from the image of the adjacent spectral continuum, 2.2859 - 2.2912 $\mu$m) shown in green. The isocontours of the integrated flux images are shown in cyan, while the isocontours of the $^{12}$CO(2-0) band image are shown in black. The position of the AGN, with an uncertainty (1$\sigma$) of $\sim 0.03$ arcsec, is marked with a black circle in the integrated flux images. The radius of these black circles represents the uncertainty (1$\sigma$) in this position.\label{fig2}}
\end{center}
\end{figure*}

The raw data of NGC 4258, together with the necessary calibrations (flat-field, dark flat-field, Ronchi-flat and arc lamp), were retrieved from the Gemini Science Archive. The programme of the observations is GN-2007A-Q-25 (PI: Storchi-Bergmann) and the observing date is 2007 April 30. The observations were taken with NIFS, in association with the adaptive optics module ALTAIR, mounted at the Gemini-north telescope. 10 exposures of the central region of NGC 4258, plus 4 sky exposures, with 600 s each, were made in the \textit{K} band, with a central wavelength of 2.2 $\mu$m. Five 15 s exposures of the A0V standard star HIP 54981 were also made. 

\begin{figure*}
\begin{center}
  \includegraphics[scale=0.41]{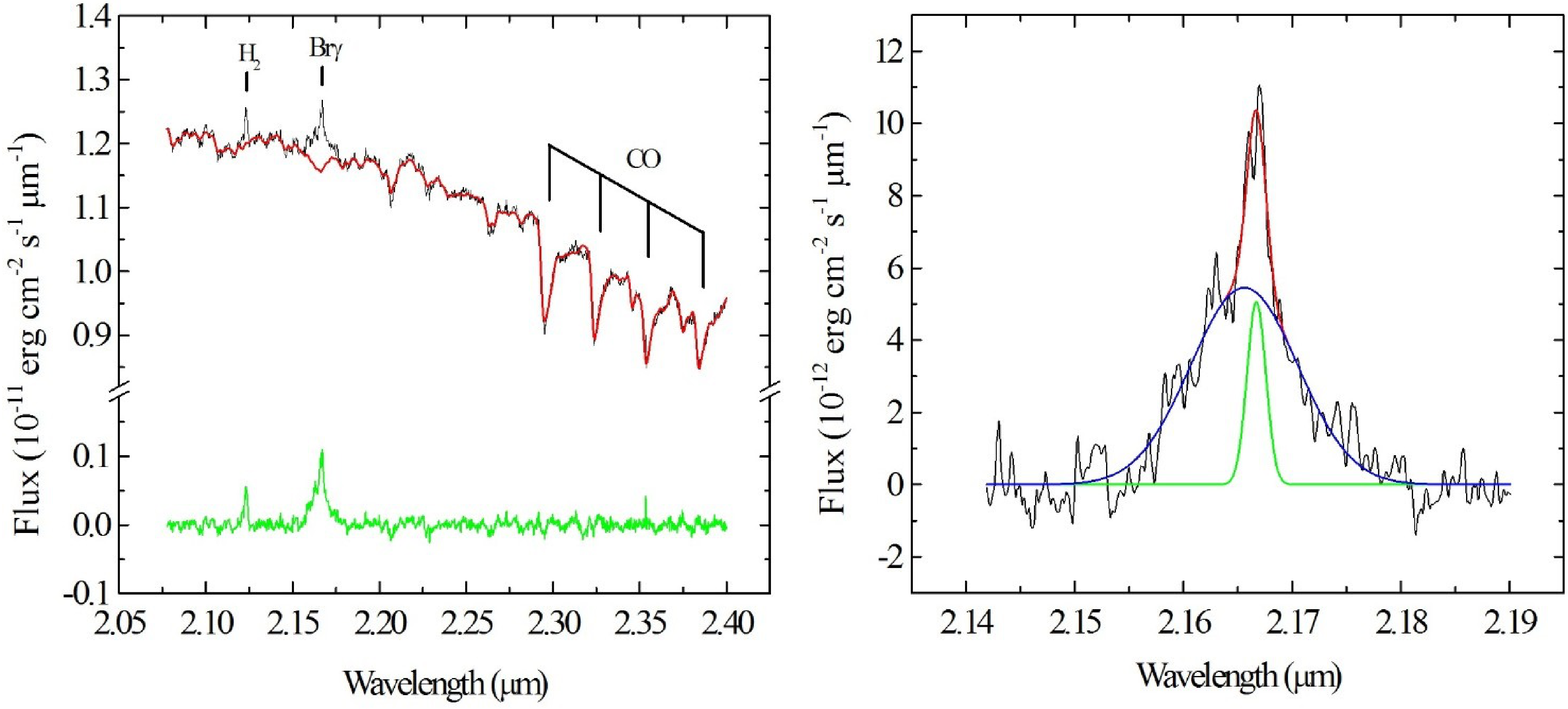}
  \caption{Left: spectrum extracted from a circular region, with a radius of 0.2 arcsec, centred on the nucleus of the galaxy (which is coincident with the position of the AGN). The fit provided by PPXF is shown in red and the fit residuals are shown in green. Right: Gaussian fits applied to the Br$\gamma$ emission line detected in the fit residuals shown at left. The green and blue curves represent the narrow and broad components, respectively, and the red curve corresponds to the final fit. Note that the blue and red curves are coincident in many areas, indicating that the narrow component is not required to reproduce these parts of the Br$\gamma$ profile.\label{fig3}}
\end{center}
\end{figure*}

The data reduction was made in \textsc{IRAF} environment, using the \textsc{Gemini} package. First, the data were trimmed, sky subtracted and bad pixel corrected. Then, a correction of gain variations was performed, using the flat-field images, and a spatial rectification was also applied, using the Ronchi-flat images. After that, the data were wavelength and flux calibrated and a telluric absorption removal was performed, using as a template the normalized median of the spectra of the observed standard star. Finally, the data cubes were constructed, with spatial pixels (spaxels) of 0.05 arcsec. The field of view (FOV) of the reduced data cubes is 3.0 arcsec $\times$ 3.0 arcsec and its spectral coverage is 2.075 - 2.425 $\mu$m, with a spectral resolution (FWHM) of 4.2 $\AA$ at 2.16 $\mu$m. Using the image corresponding to the blue wing of the broad component of the Br$\gamma$ emission line (2.1564 - 2.1608 $\mu$m; see Section 3), we obtained a value of 0.22 arcsec for the FWHM of the point spread function (PSF) of the observations.

The reduced data cubes were treated with a procedure including: correction of the differential atmospheric refraction \citep{fil82}, calculation of the median of the data cubes, spatial re-sampling followed by an interpolation, Butterworth spatial filtering \citep{gon02} and `instrumental fingerprint' removal. This entire process was performed using scripts written in \textsc{Interactive Data Language} (\textsc{IDL}) and is described in further detail in \citet{men14} and \citet{men15}. The differential atmospheric refraction generates a displacement of the spatial structures along the spectral axis of the data cube. Although this effect is weak in the infrared, we verified that the correction procedure results in visible improvements in NIFS data cubes. A median of the resulting data cubes was calculated to correct for remaining bad pixels and cosmic rays, not removed in the data reduction. After that, a spatial resampling, together with an interpolation of the values, was applied in order to obtain spaxels of 0.021 arcsec. This step of the data treatment does not change the spatial resolution of the observation, but improves the visualization of the contours of the structures. A Butterworth spatial filtering was then applied to each image of the resampled data cubes to remove high spatial frequency components. The last step of the treatment procedure involved the use of the Principal Component Analysis (PCA) Tomography technique \citep{ste09} to isolate and remove the instrumental fingerprint, which has a characteristic low-frequency spectral signature and appears as vertical stripes across the images of the data cube. Although the origin of the NIFS instrumental fingerprint is unknown, our approach involving PCA Tomography has proved to be very effective in removing this instrumental feature.

The integrated flux image and the mean spectrum, over the entire FOV, of the treated data cube are shown in Fig.~\ref{fig1}. Prominent CO absorption bands and a weak H$_2 \lambda 2.1218$ $\mu$m emission line (corresponding to the ro-vibrational transition 1-0S(1)) can be seen in the mean spectrum. The integrated flux image reveals a compact nuclear emission.

\section{Analysis of the emission line spectrum}

The first goal of this study is to analyse the emission line spectrum of the data cube of NGC 4258. Using scripts written in \textsc{IDL}, we passed the spectra of the treated data cube to the rest frame, assuming a redshift of \textit{z} = 0.001494 (NASA Extragalactic Database - NED). We also sampled the redshift-corrected spectra with $\Delta\lambda = 1\AA$. After that, in order to perform the starlight subtraction of the data, we applied the Penalized Pixel Fitting (PPXF) \citep{cap04}, which combines template spectra from a given base, convolved with a Gauss-Hermite expansion, to reproduce the stellar spectrum of an object. The convolution with a Gauss-Hermite expansion has the purpose of reproducing the widths, shifts (due to the radial velocity of the object) and also asymmetries of the observed absorption lines. The base used in this work is composed of stellar spectra observed with NIFS and is described in \citet{win09}. The PPXF method was applied to the spectrum corresponding to each spaxel of the data cube. We subtracted the synthetic stellar spectra provided by the fitting procedure from the observed ones, and obtained a data cube with only emission lines.

\begin{figure*}
\begin{center}
  \includegraphics[scale=0.357]{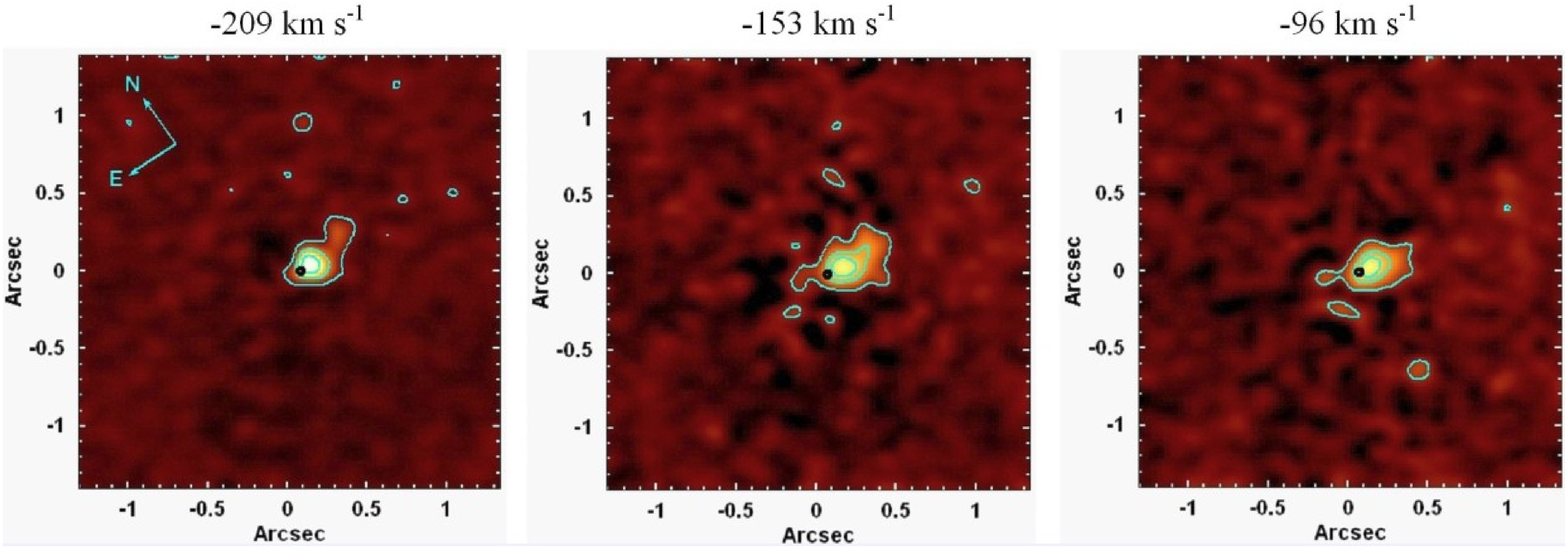}
  \includegraphics[scale=0.33]{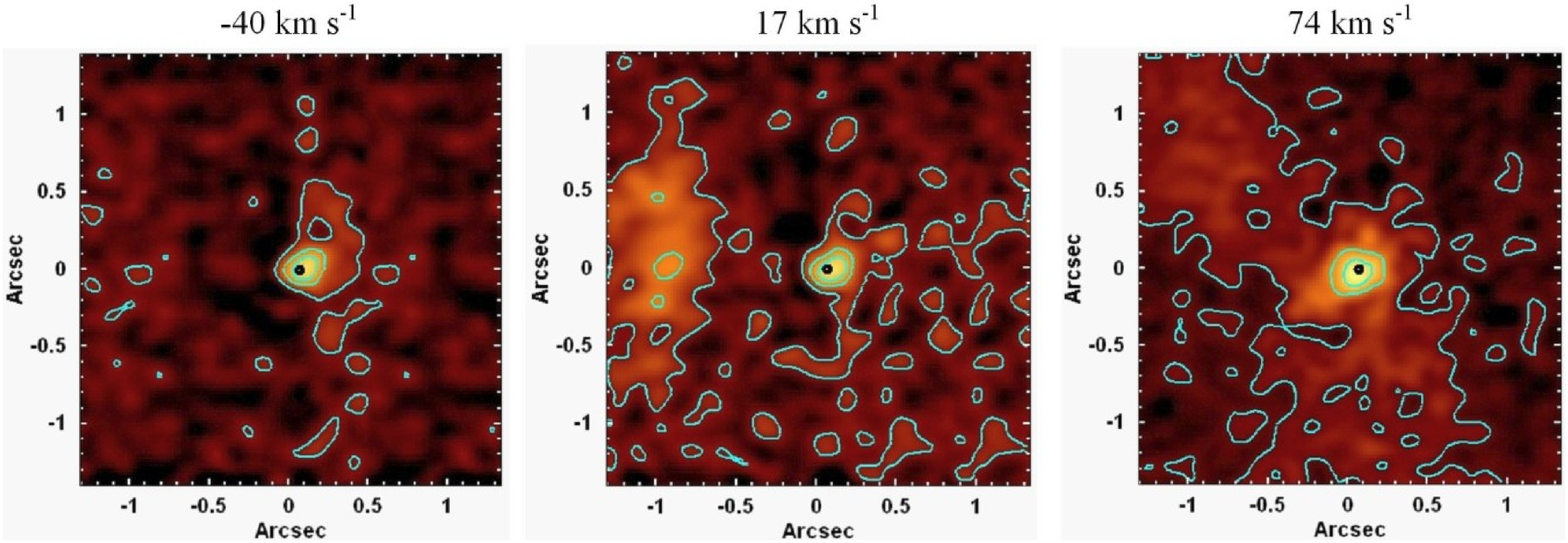}
  \includegraphics[scale=0.33]{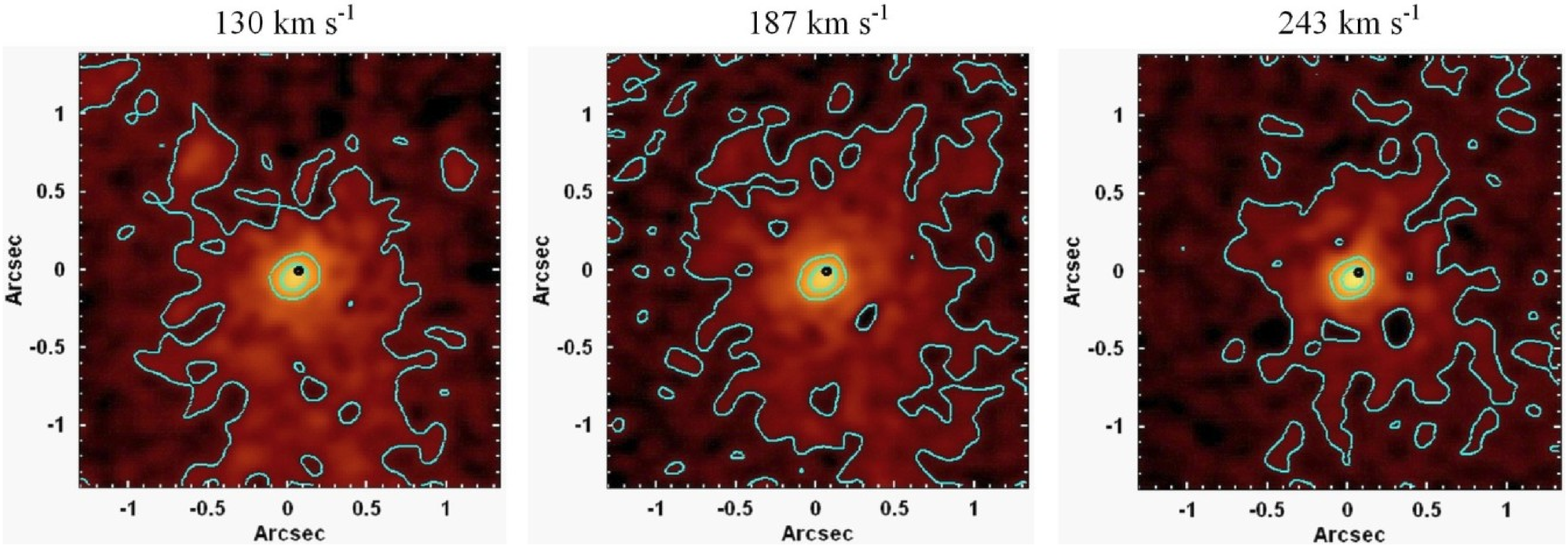}
  \includegraphics[scale=0.33]{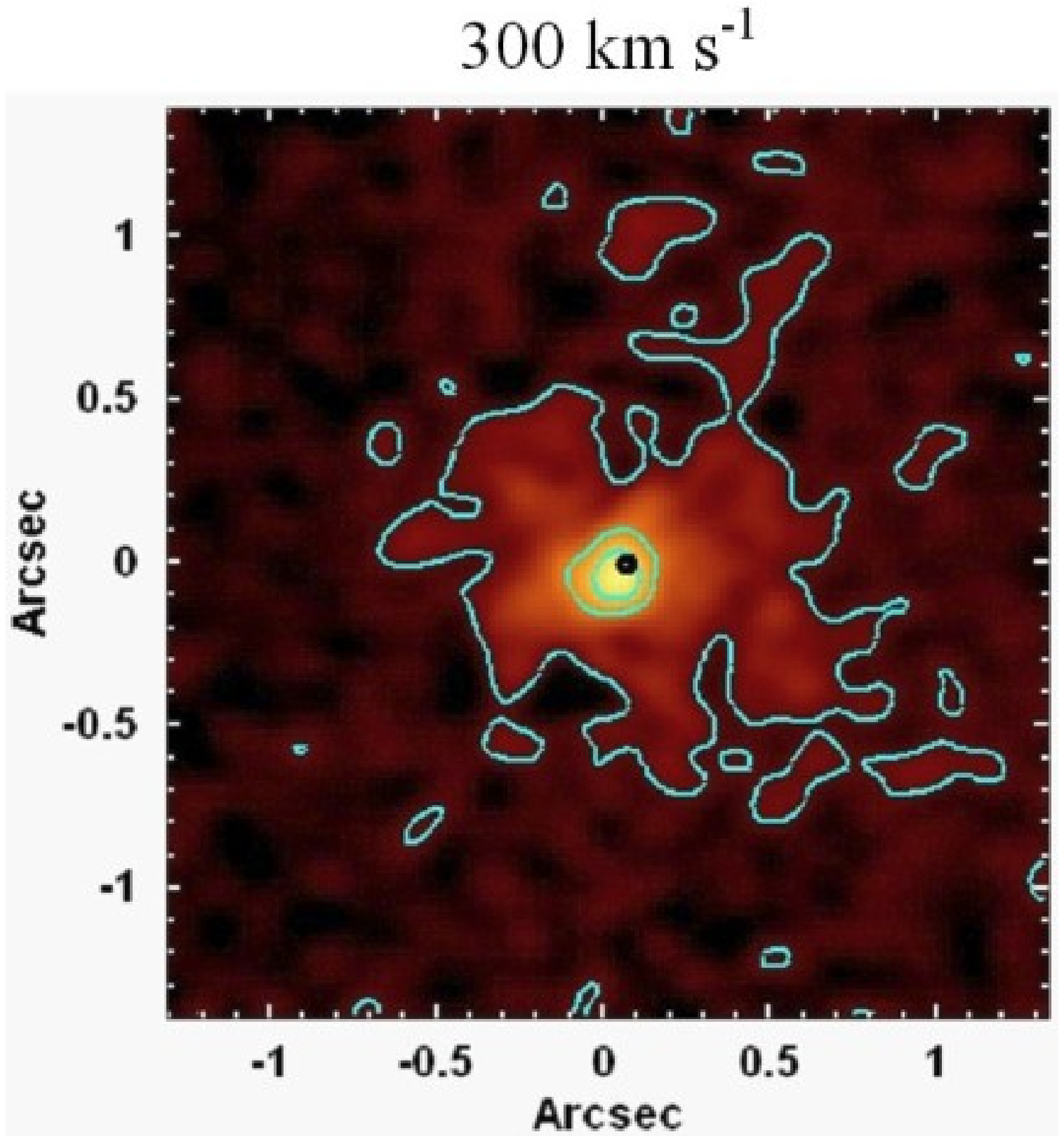}
  \caption{Channel maps of the H$_2 \lambda 2.1218$ $\mu$m emission line. Each map was constructed with a velocity range of $\sim 40$ km s$^{-1}$. The numbers above the maps correspond to the average velocities of the intervals used to construct the maps. The position of the AGN is marked with a black circle.\label{fig4}}
\end{center}
\end{figure*}

Only two emission lines are visible in the starlight subtracted data cube: H$_2 \lambda 2.1218$ $\mu$m and Br$\gamma$, the latter having an apparent broad component. Fig.~\ref{fig2} shows integrated flux images of the two detected emission lines, with isocontours, and the amplitude/noise (A/N) ratio maps of these lines. The noise values, required for the calculation of the A/N ratios, were taken as being equal to the standard deviation of the values of the starlight subtracted spectra in spectral regions adjacent to the emission lines. In order to evaluate the exact position of the AGN in NGC 4258, we constructed a red-green (RG) composite, with reduced FOV, with the image of the blue wing ($2.1564 \mu$m - $2.1608 \mu$m) of the broad component of Br$\gamma$ in red and the image of the $^{12}$CO(2-0) band (obtained by subtracting the image corresponding to the wavelength interval of the $^{12}$CO(2-0) band, 2.2929 $\mu$m - 2.2982 $\mu$m, from the image of the adjacent spectral continuum, 2.2859 $\mu$m - 2.2912 $\mu$m) in green. The result, with the isocontours of the $^{12}$CO(2-0) band image, is also shown in Fig.~\ref{fig2}. The image of the broad wing of Br$\gamma$ indicates the position of the AGN, while the $^{12}$CO(2-0) band image represents the central part of the stellar bulge. So, based on the RG composite in Fig.~\ref{fig2}, we conclude that the position of the AGN is essentially coincident with the position of the stellar nucleus, as expected. The integrated flux images in Fig.~\ref{fig2} reveal that the Br$\gamma$ and H$_2 \lambda 2.1218$ $\mu$m emission comes from a compact region centred on the position of the AGN (coincident with the nucleus) of the galaxy. However, the area corresponding to the H$_2 \lambda 2.1218$ $\mu$m emission is more elongated along the east-west direction. The A/N ratio of both emission lines is lower than 10 in most of the FOV, with maximum values of $\sim 23$ and $\sim 14$ for Br$\gamma$ and H$_2 \lambda 2.1218$ $\mu$m, respectively. The A/N ratio map of the H$_2 \lambda 2.1218$ $\mu$m line also shows a diffuse emission northeast from the nucleus. A careful inspection of the starlight subtracted data cube confirms that Br$\gamma$ is only visible in the nuclear compact emitting area, while H$_2 \lambda 2.1218$ $\mu$m is visible in the nuclear emitting region and in the diffuse area northeast from the nucleus. 

In order to analyse the nuclear emission in further detail, we extracted the spectrum of a circular region, centred on the nucleus, with a radius of $0\arcsec\!\!.2$, from the data cube before the starlight subtraction. The extracted spectrum is shown in Fig.~\ref{fig3}, together with the fit provided by PPXF and the fit residuals. It is easy to see that the synthetic spectrum obtained with PPXF reproduces the continuum very well and also the absorption features in the observed spectrum. The H$_2 \lambda 2.1218$ $\mu$m and Br$\gamma$ emission lines are very prominent in the fit residuals. The profile of Br$\gamma$ clearly suggests the presence of a broad component. In order to evaluate that, we fitted the Br$\gamma$ emission line in the spectrum in Fig.~\ref{fig3} with a sum of two Gaussian functions with different velocities and widths. The fit, also shown in Fig.~\ref{fig3}, reproduces, with good precision, the main properties of the profile of Br$\gamma$. The values of the FWHM, corrected for the instrumental spectral resolution, of the Gaussians representing the narrow and broad components of Br$\gamma$ are FWHM$_{Br\gamma}(narrow) = 260 \pm 15$ km s$^{-1}$ and FWHM$_{Br\gamma}(broad) = 1600 \pm 29$ km s$^{-1}$, respectively. The broad component has a velocity of $V_{Br\gamma}(broad) = -156 \pm 14$ km s$^{-1}$ relative to the narrow component. The integrated fluxes of the broad and narrow components of Br$\gamma$ are $F_{Br\gamma}(broad) = (6.73 \pm 0.27) \times 10^{-15}$ erg cm$^{-2}$ s$^{-1}$ and $F_{Br\gamma}(narrow) = (1.11 \pm 0.09) \times 10^{-15}$ erg cm$^{-2}$ s$^{-1}$, respectively. The integrated flux of the H$_2 \lambda 2.1218$ $\mu$m line in the same spectrum is $F_{H_2} = (1.31 \pm 0.04) \times 10^{-15}$ erg cm$^{-2}$ s$^{-1}$. 

\begin{figure}
\begin{center}
  \includegraphics[scale=0.24]{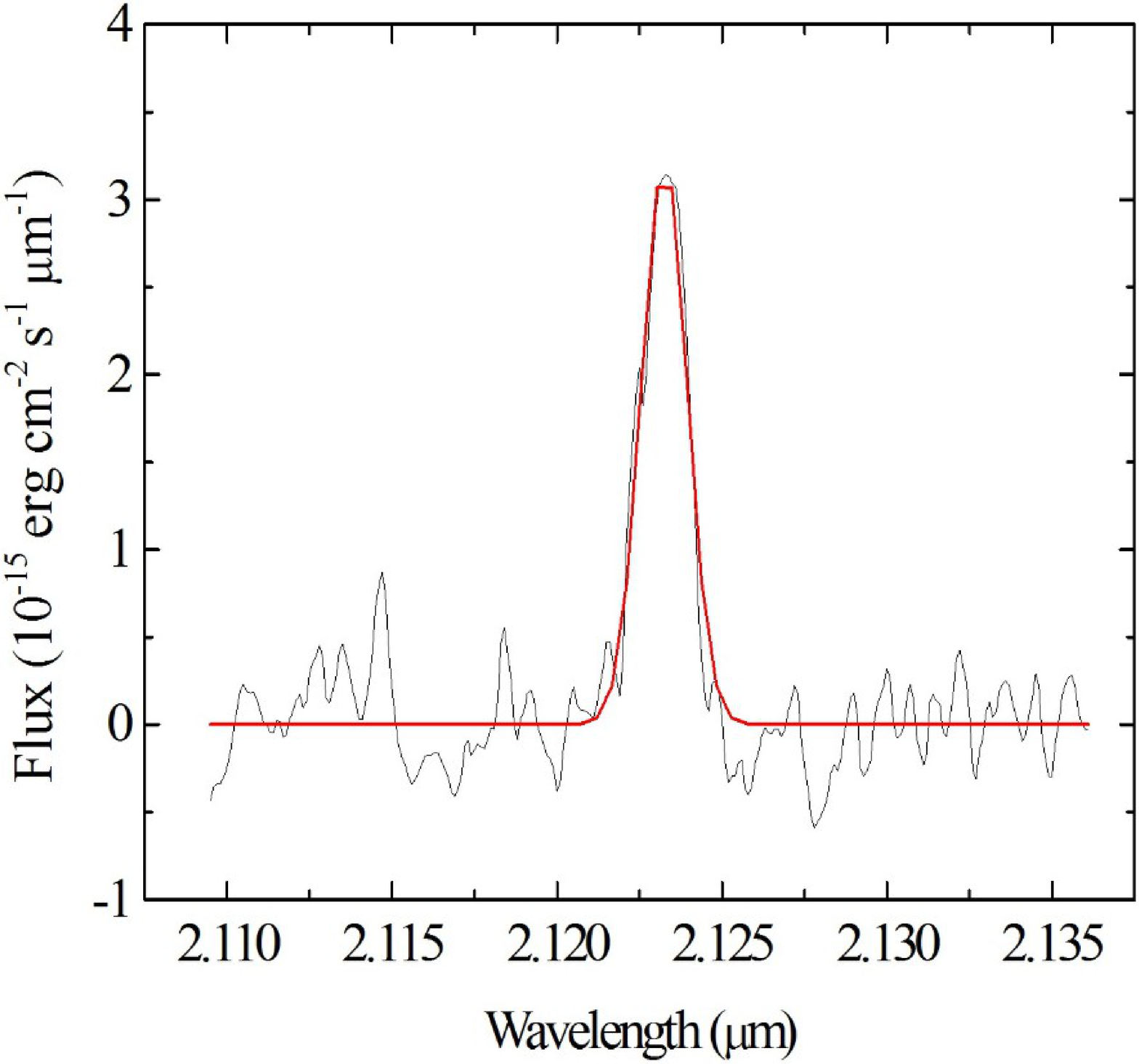}
  \caption{Example of a Gaussian function fitted to the H$_2 \lambda 2.1218$ $\mu$m line of a spectrum in the nuclear emitting region of the data cube of NGC 4258.\label{fig5}}
\end{center}
\end{figure}

It is well known that diagnostic diagrams with ratios of the H$_2$ lines can be used to determine whether thermal or non-thermal processes are responsible for the molecular emission \citep{mou94}. However, we could not perform such analysis here, as we detected only the H$_2 \lambda 2.1218$ $\mu$m emission line in the data cube of NGC 4258. A diagnostic diagram of [Fe \textsc{II}] $\lambda 1.4$ $\mu$m/Br$\gamma$ $\times$ H$_2 \lambda 2.1218$ $\mu$m/Br$\gamma$ can also be used to determine the dominant excitation mechanism \citep{mai17} but, since we do not know the flux of the [Fe \textsc{II}] $\lambda 1.64$ $\mu$m line in the nuclear emitting region, this analysis was not possible as well. On the other hand, the H$_2 \lambda 2.1218$ $\mu$m/Br$\gamma$ ratio alone may provide, at least, limited information about this topic. So, we calculated the H$_2 \lambda 2.1218$ $\mu$m/Br$\gamma$ ratio for the nuclear spectrum, after the starlight subtraction, shown in Fig.~\ref{fig3}. Only the narrow component of the Br$\gamma$ line, obtained with the Gaussian fitting described before, was taken into account. The obtained value, H$_2 \lambda 2.1218$ $\mu$m/Br$\gamma$ = $1.19 \pm 0.11$, indicates excitation by shocks or by an AGN, according to the diagram in Fig. 2 of \citet{mai17}. Considering that this galaxy has a known AGN, we conclude that AGN excitation is the most likely mechanism responsible for the molecular H$_2$ emission detected in this work.

\begin{figure*}
\begin{center}
  \includegraphics[scale=0.55]{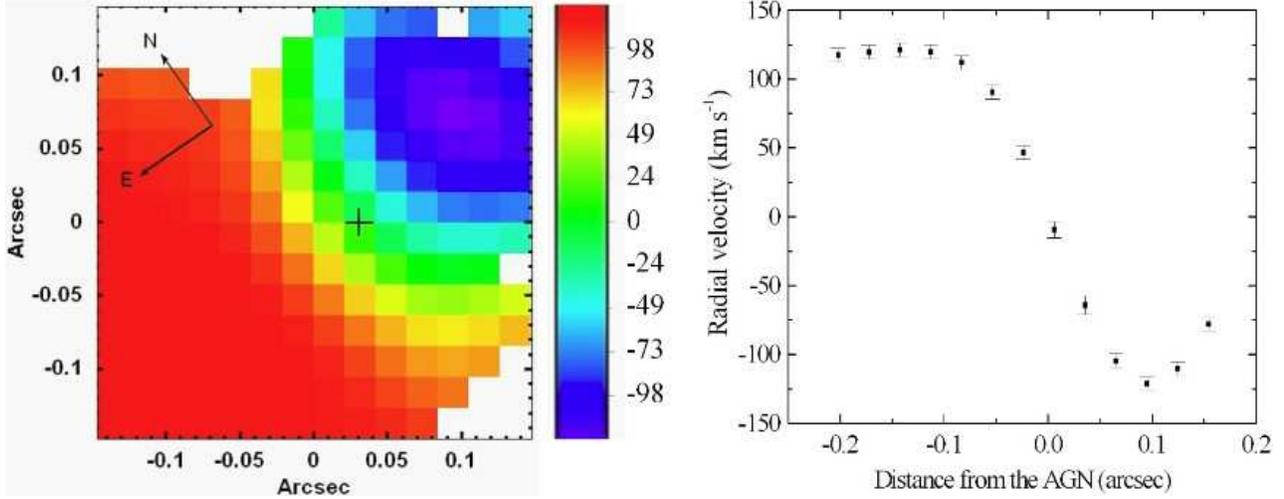}
  \caption{Left: $V_{H_2}$ map of the data cube of NGC 4258, with reduced FOV. The $V_{H_2}$ values are in km s$^{-1}$. Areas with A/N ratios of the H$_2 \lambda 2.1218$ $\mu$m line lower than 6 were masked. The black cross corresponds to the position of the AGN. Right: $V_{H_2}$ curve extracted along the line of nodes of the map shown at left.\label{fig6}}
\end{center}
\end{figure*}

We detected no kinematics associated with the Br$\gamma$ emission line. On the other hand, although the H$_2 \lambda 2.1218$ $\mu$m emitting area is very compact, this line shows a kinematic pattern that deserves special attention. We constructed channel maps of H$_2 \lambda 2.1218$ $\mu$m, which are shown in Fig.~\ref{fig4}. The molecular emission with radial velocities between -209 and -96 km s$^{-1}$ is located west from the AGN, while the molecular emission with radial velocities between 130 and 300 km s$^{-1}$ is located east from the AGN. This kinematic behaviour takes place along $PA_{H_2} = 91\degr \pm 5\degr$. We verified that, despite the irregularities and asymmetries of the H$_2 \lambda 2.1218$ $\mu$m line, Gaussian fits provide reliable values for the radial velocity ($V_{H_2}$) of this line (with errors between 5 and 10 km s$^{-1}$) for A/N ratios higher than 6. Fig.~\ref{fig5} shows an example of a Gaussian function fitted to the H$_2 \lambda 2.1218$ $\mu$m line of a spectrum in the nuclear emitting region of the data cube. We can see that, although the profile of the line is not exactly reproduced by the Gaussian function, the fit provides an accurate estimate of the wavelength (and, therefore, of the radial velocity) corresponding to the peak of the line. We constructed a $V_{H_2}$ map, using the $V_{H_2}$ values obtained with Gaussian fits of this line, but only taking into account the FOV within a radius of $\sim 0\arcsec\!\!.2$ from the nucleus. The reason for this reduced FOV is that the radial velocity values in other areas of the FOV were not reliable (due to the low A/N ratios and the irregularities of the line). We assumed that $V_{H_2} = 0$ km s$^{-1}$ at the position of the AGN. In other words, the radial velocities were obtained relative to the AGN. The $V_{H_2}$ map, shown in Fig.~\ref{fig6}, reveals a morphology consistent with a rotating disc around the nucleus. The PA of the line of nodes of the $V_{H_2}$ map is $93\degr \pm 6\degr$, which is compatible, at 1$\sigma$ level, with the PA resulting from the analysis of the channel maps. We extracted a curve, from the $V_{H_2}$ map, along the axis corresponding to the line of nodes. The curve, shown in Fig.~\ref{fig6}, has an amplitude of 120 km s$^{-1}$ and shows a pattern indicative of rotation around the nucleus. In addition, an asymmetry between the regions with positive and negative values can also be noted. This may indicate that the corresponding molecular disc is eccentric. A velocity dispersion map of the $H_2 \lambda 2.1218$ $\mu$m line could help to evaluate this hypothesis; however, due to the asymmetries and irregularities of this line, the velocity dispersion values provided by the Gaussian fits were not reliable enough to be analysed in this work. A dynamical modelling of this molecular disc could also help to evaluate this hypothesis and could provide important parameters such as the inclination of the disc. Unfortunately, the number of resolution elements within the area corresponding to the molecular disc is very small, which makes any dynamical modelling too uncertain to be included in this work.

The analysis of the H$_2$ kinematics in other areas of the FOV is more complicated, as the H$_2 \lambda 2.1218$ $\mu$m line was not clearly detected there (with the exception of the diffuse area northeast from the nucleus). Nevertheless, we performed a spatial binning in the data cube, in order to increase the size of the spaxels and also the A/N ratios of the H$_2 \lambda 2.1218$ $\mu$m line. However, even with spaxels of 0.3 arcsec, the A/N ratios of this line remained lower than 6 in most of the FOV. As a consequence, we were not able to perform a reliable analysis of the H$_2$ kinematics in areas located farther from the nucleus.

\begin{figure*}
\begin{center}
  \includegraphics[scale=0.57]{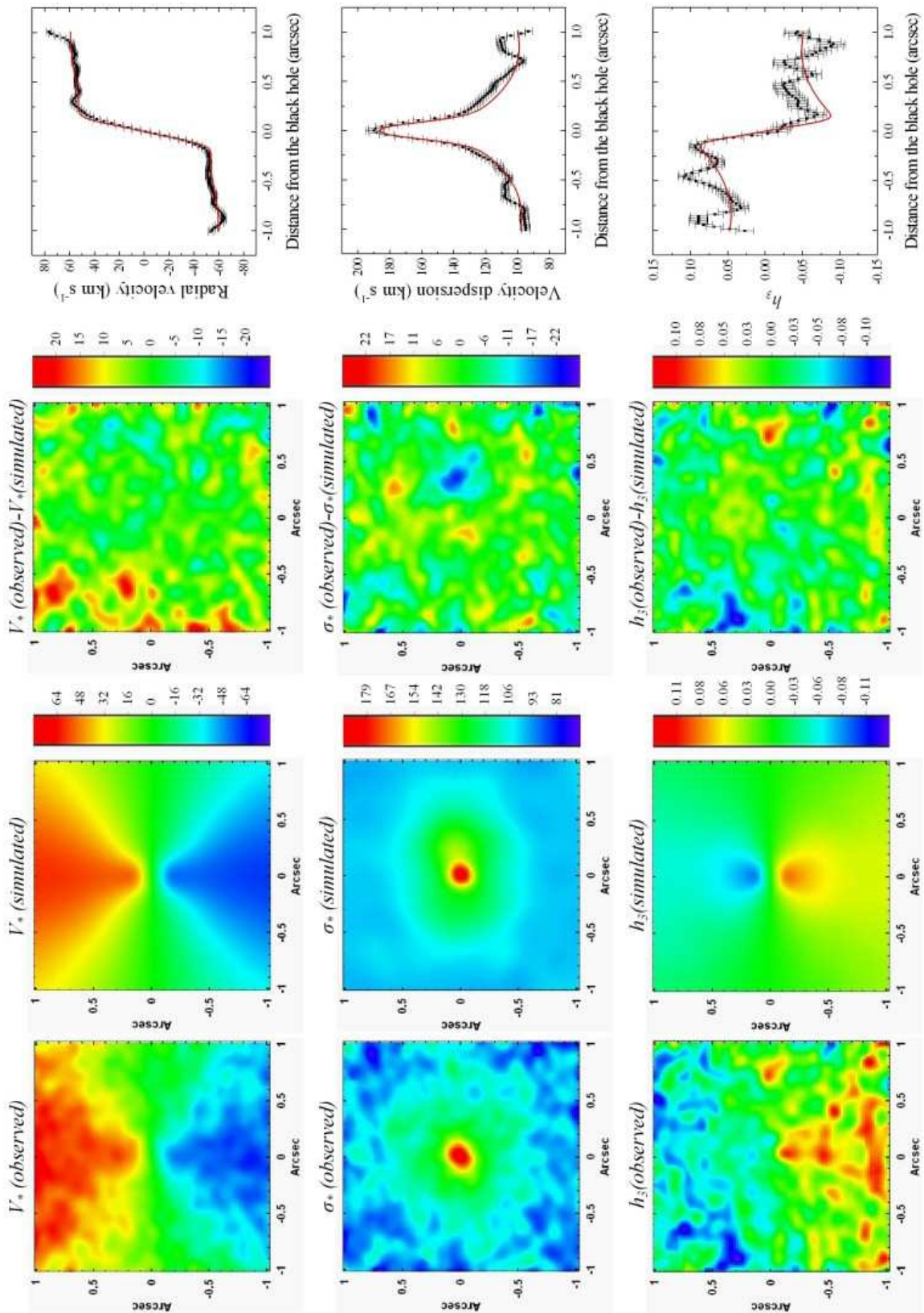}
  \caption{Observed and simulated stellar kinematic maps of the data cube of NGC 4258. The simulated maps were obtained with a model of a thin rotating stellar circular disc around the SMBH. The residual maps reveal the good agreement between the observed and simulated parameters. The curves extracted from the observed maps, along the axis corresponding to the line of nodes of the $V_*$ map, are also shown. Superposed to these observed curves are the simulated ones, shown in red, which were extracted from the simulated maps, along the same axis.\label{fig7}}
\end{center}
\end{figure*}

\section{Analysis of the stellar disc}

As already discussed before, since NGC 4258 has the SMBH with one of the most precise mass determinations, it is the ideal target to test the reliability of different methods used to estimate the SMBH mass. So, we applied the model of a thin rotating stellar disc, superimposed to a bulge component, to our data cube of NGC 4258, in order to reproduce the observed stellar kinematic parameters and to evaluate the precision of the SMBH mass obtained with this procedure. This model was chosen due to its simplicity and compatibility with the observed stellar radial velocity map ($V_*$ map, see Fig.~\ref{fig7}), which certainly suggests the presence of a rotating disc. It is important to emphasize, however, that our purpose here is not to try to obtain necessarily a more precise estimate for the SMBH mass in NGC 4258 than those of previous works. Our main goal here is only to test the reliability of our method, which, despite its simplicity, has the advantage of not requiring axysimmetry and being applicable to asymmetric nuclear regions (with eccentric discs, for example). See more details in Section 5.

Besides a synthetic stellar spectrum, the PPXF method (discussed in Section 3) also provides, for an observed spectrum, the values of $V_*$, the stellar velocity dispersion ($\sigma_*$) and the Gauss-Hermite coefficients $h_3$ and $h_4$. Since this procedure was applied to the spectrum corresponding to each spaxel of the data cube of NGC 4258, we obtained maps of these kinematic parameters. We also extracted curves of these maps along an axis corresponding to the line of nodes of the $V_*$ map. All these results are shown in Fig.~\ref{fig7}. The only exceptions are the map and curve of $h_4$, which are quite noisy, do not show clear structures and, because of that, were not included in this study.  Since we are more interested in the stellar kinematics close to the SMBH and also considering the lower signal-to-noise (S/N) ratio values in the most peripheral areas of the FOV, we decided to analyse only the stellar kinematics in the inner 2 arcsec$^2$. We assumed that the centre of the $V_*$ map, with $V_* = 0$ km s$^{-1}$, corresponds to the position of the nucleus. Therefore, the $V_*$ map in Fig.~\ref{fig7} shows the radial velocities relative to the AGN. The uncertainties of the values were obtained by a Monte Carlo procedure. We constructed, for each spectrum, a histogram of the values in certain areas without emission lines. This process was applied to the starlight subtracted spectra. After that, Gaussian functions were fitted to these histograms and Gaussian distributions of random numbers (random noise), with the same standard deviations of the Gaussians fitted to the histograms, were generated. These distributions of random noise were sequentially added to the synthetic stellar spectra corresponding to the fits obtained with the PPXF. In other words, for each spaxel of the data cube, we created a number of spectra with random noise, corresponding to the sum of the synthetic stellar spectrum provided by the PPXF and random noise distributions. Finally, the PPXF was applied to all the spectra with random noise and the uncertainty of a given kinematic parameter for each spaxel was taken as the standard deviation of the values obtained with all these PPXF fits.

\begin{figure}
\begin{center}
  \includegraphics[scale=0.40]{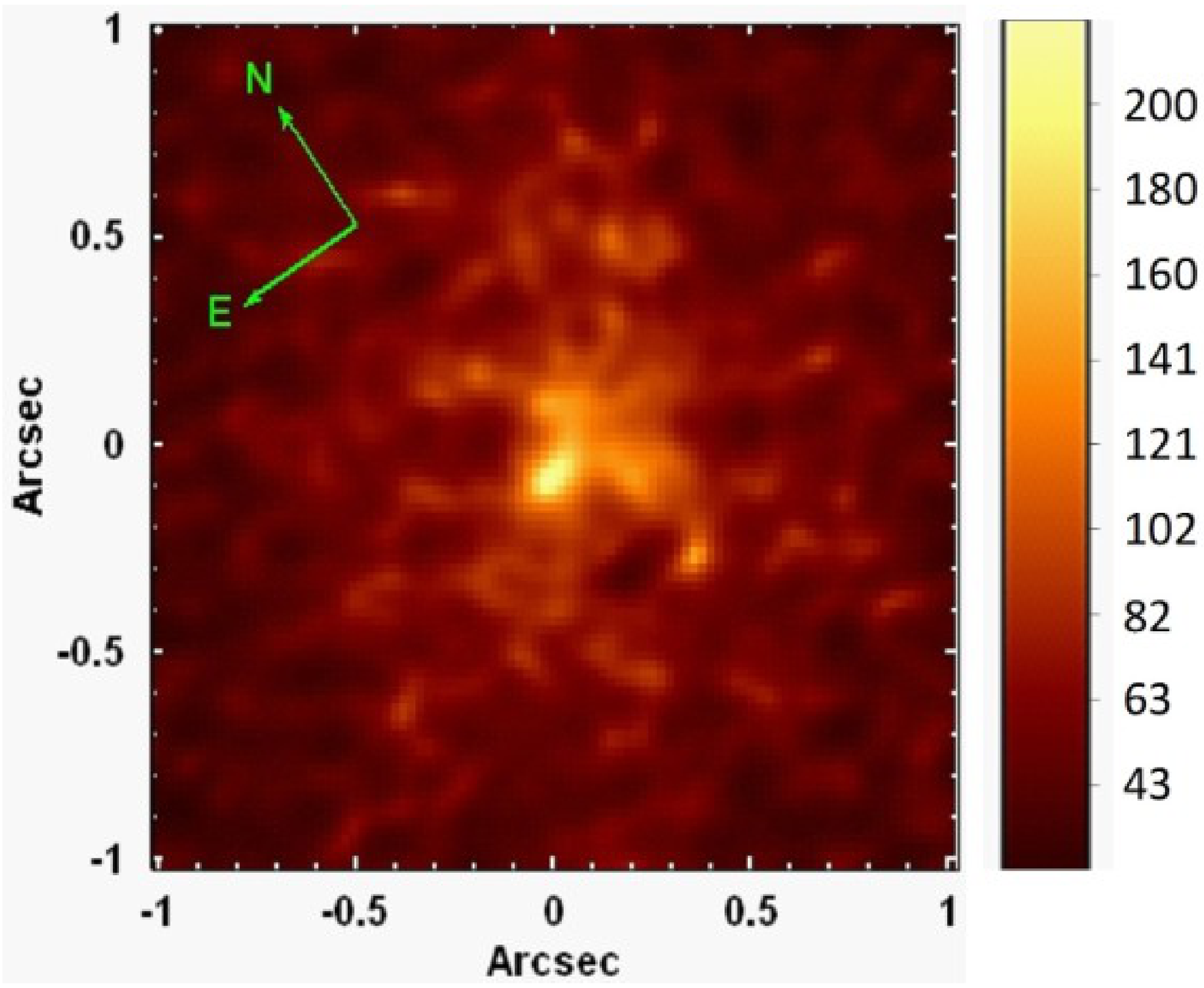}
  \caption{S/N ratio map of the wavelength interval 2.236 - 2.259 $\mu$m, with the same FOV of the data cube used to perform the dynamical modelling.\label{fig8}}
\end{center}
\end{figure}

The map and curve of $V_*$ show a symmetric rotation pattern around the nucleus, with velocity moduli lower than 70 km s$^{-1}$. The PA of the line of nodes is $PA_{V_*} = -35\degr \pm 5\degr$. The map and curve of $\sigma_*$ reveal a peak in a position coincident with the nucleus, with a value of $\sigma_*(peak) = 190$ km s$^{-1}$. The $h_3$ map is considerably noisier than the previous two, but we can see an anticorrelation between the maps and curves of $V_*$ and $h_3$. The latter show values with moduli lower than 0.14. Fig.~\ref{fig8} shows a S/N ratio map of the wavelength interval 2.236 - 2.259 $\mu$m. The values are higher than 30 in most of the FOV, and even higher than 200 close to the nucleus, indicating that the structures observed in the maps of the kinematic parameters are reliable. As expected, all these results are consistent with those obtained by \citet{dre15}, who analysed the stellar kinematics of the same data cube used here and applied a dynamical modelling using the Jeans method.

The model we used to reproduce the maps of $V_*$, $\sigma_*$ and $h_3$ assumes a thin rotating stellar disc around the SMBH, located at the position of the nucleus. The fact that the $V_*$ map is symmetric and that $\sigma_*(peak)$ is coincident with the nucleus suggests that such rotating disc is circular, so only circular orbits were considered in the model. The same model was used in \citet{men15b}; however, in that case, the modelled stellar disc was eccentric, as some of the observed features, such as an asymmetric $V_*$ map and an off-centred $\sigma_*(peak)$, could not be reproduced by a model of a circular disc. The main steps of this dynamical modelling are as follows.

\begin{figure*}
\begin{center}
  \includegraphics[scale=0.50]{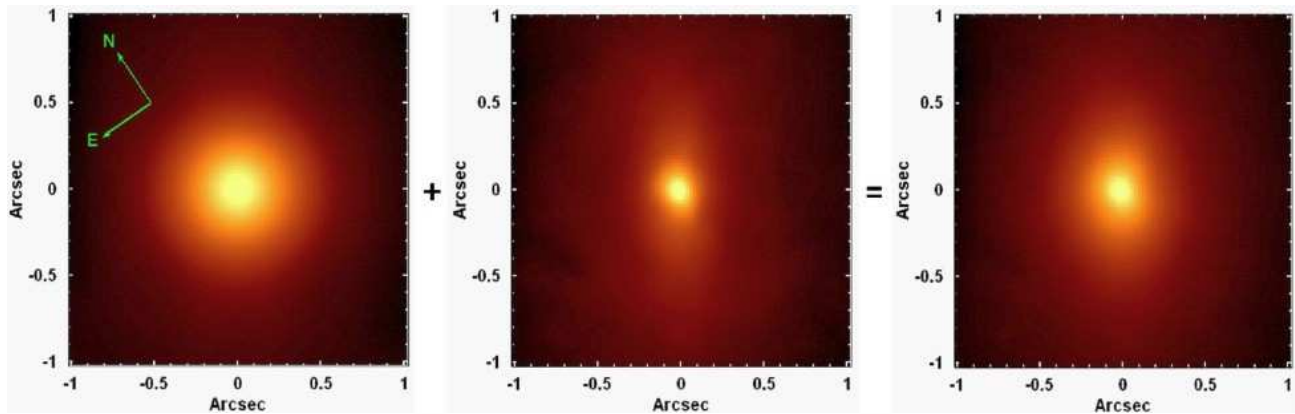}
  \caption{Left: symmetric component (representing the central part of the stellar bulge) of the \textit{I}-band \textit{HST} image of NGC 4258, corresponding to a S\'ersic profile with an index of 2.0. Centre: asymmetric component (representing the stellar disc) of the \textit{I}-band \textit{HST} image of NGC 4258, obtained by subtracting the symmetric component from the original image. Right: original \textit{I}-band \textit{HST} image of NGC 4258. All the images have the same FOV of the NIFS data cube used to perform the dynamical modelling.\label{fig9}}
\end{center}
\end{figure*}

\begin{itemize}

\item Superimposition of 122 concentric circular orbits, taking as free parameters the mass of the SMBH ($M_{\bullet}(disc)$) and the inclination of the disc ($i_{disc}$). The stellar mass within each orbit was calculated using an \textit{HST} image of this galaxy (retrieved from the \textit{HST} data archive), obtained with the Wide Field Camera 3 in the \textit{F814W} filter (\textit{I} band). This image was decomposed into a symmetric component, which corresponds to the central part of the stellar bulge, and an asymmetric one, corresponding to the stellar disc. The best decomposition was obtained assuming a symmetric component given by a S\'ersic profile, with a S\'ersic index of 2.0. The original \textit{HST} image and the result of this decomposition are shown in Fig.~\ref{fig9}. The symmetric component was deprojected, using a Multi-Gaussian Expansion \citep{cap02}, which resulted in a luminosity density profile (in $L_{I\sun}/pc^3$). With all these data, the stellar mass was determined assuming a mass-to-light ratio in the \textit{I} band ($M/L_I$), which is another free parameter of the model.

\item Projection of the velocities obtained for all the concentric orbits on the plane of the sky, which resulted in a synthetic velocity map.

\item Construction of two synthetic data cubes. The first represents the central part of the stellar bulge, while the second represents the stellar disc. Both data cubes were constructed using a representative spectrum, in the range of 2.255 - 2.389 $\mu$m, given by the fit provided by PPXF applied to the mean spectrum of the data cube of NGC 4258, with the template spectra not convolved with a Gauss-Hermite expansion. The purpose of fitting the mean spectrum of the data cube, without convolving the template spectra with the Gauss-Hermite expansion, is to obtain a representative stellar spectrum with no broadening or asymmetries in the absorption lines. The data cube corresponding to the stellar disc contains the representative spectrum shifted according to the radial velocity values of the synthetic velocity map obtained above. On the other hand, all the spaxels of the data cube corresponding to the central part of the bulge contain the representative spectrum, with a radial velocity equal to 0 km s$^{-1}$. The spectra of the disc and bulge synthetic data cubes were convolved with estimates of the stellar velocity dispersions of the disc ($\sigma_d$) and of the bulge ($\sigma_b$), respectively, which are also free parameters in this model.

\item Combination of the bulge and disc synthetic data cubes. This procedure was performed using the symmetric and asymmetric components obtained from the decomposition of the \textit{HST} image to determine the relative weights between the spaxels of the two synthetic data cubes.

\item Convolution of all images of the combined synthetic data cube with an estimate of the PSF of the observation.

\item Application of the PPXF method to the synthetic data cube, in order to obtain the simulated maps of the kinematic parameters.

\end{itemize}

The free parameters of our model are $M_{\bullet}(disc)$, $i_{disc}$, $M/L_I$, $\sigma_d$ and $\sigma_b$. Such parameters were varied and the procedure mentioned above was repeated, in order to minimize the total $\chi^2$ ($\chi_T^2$; Menezes \& Steiner 2015), given by
\\
\begin{equation}
\resizebox{.98\hsize}{!}{$\chi^2_T=\frac{(\frac{1}{Min(\chi^2_{V_*})})*\chi^2_{V_*}+(\frac{1}{Min(\chi^2_{\sigma_*})})*\chi^2_{\sigma_*}+(\frac{1}{Min(\chi^2_{h_3})})*\chi^2_{h_3}}{(\frac{1}{Min(\chi^2_{V_*})})+(\frac{1}{Min(\chi^2_{\sigma_*})})+(\frac{1}{Min(\chi^2_{h_3})})}$},
\end{equation}
\\
where $\chi^2_{V_*}$, $\chi^2_{\sigma_*}$ and $\chi^2_{h_3}$ are the values of the $\chi^2$ obtained for the maps of $V_*$, $\sigma_*$ and $h_3$, respectively, and $Min(\chi^2_{V_*})$, $Min(\chi^2_{\sigma_*})$ and $Min(\chi^2_{h_3})$ are the minimum values found for $\chi^2_{V_*}$, $\chi^2_{\sigma_*}$ and $\chi^2_{h_3}$, respectively.

The definition of the $\chi^2$ of a given parameter $A$ is similar to the one adopted in \citet{men15b}:
\\
\begin{equation}
\resizebox{.98\hsize}{!}{$\chi^2_{A}=\frac{1}{I}\sum_{i=1}^{N_x}\sum_{j=1}^{N_y}\frac{I_{ij}\cdot \left(A_{ij}(observed)-A_{ij}(simulated)\right)^2}{\sigma_{A_{ij}}^2}$}.
\end{equation}
\\
In the equation above, $A_{ij}(observed)$ and $A_{ij}(simulated)$ are the values of the observed and simulated maps of $A$, respectively, at the spaxel $(i,j)$, $I$ is the integrated flux of the data cube, $I_{ij}$ is the integrated flux of the spectrum corresponding to spaxel $(i,j)$ of the data cube, $\sigma_{A_{ij}}$ is the uncertainty of the parameter $A_*$ at the position corresponding to spaxel $(i,j)$, $N_x$ is the number of spaxels along the horizontal spatial axis and $N_y$ is the number of spaxels along the vertical spatial axis.

The simulated maps of $V_*$, $\sigma_*$ and $h_3$ provided by the best obtained model, together with the simulated curves extracted, from these maps, along the line of nodes, are shown in Fig.~\ref{fig7}. In order to make easier to evaluate the agreement between the observed and modelled data, we also included residual maps, corresponding to the difference between the observed and simulated kinematic maps, in Fig.~\ref{fig7}. Most of the simulated $V_*$ curve is compatible with the observed $V_*$ curve, at $1\sigma$ or $2\sigma$ levels. The residual map of $V_*$ reveals certain discrepancies in areas north and east from the nucleus, where the simulated values are significantly lower than the observed ones. It is not clear whether these patterns in the residual map are due to limitations of the model or due to irregularities in the observed $V_*$ map, possibly related to imprecisions of the PPXF method. Despite these discrepancies, however, the model reproduced very well the observed $V_*$ map. The main properties of the observed $\sigma_*$ map were also well reproduced by the model. Again, most of the simulated $\sigma_*$ curve is compatible with the observed points, at $1\sigma$ or $2\sigma$ levels. The main discrepancy, in this case, occurs in a small area $\sim 0.35$ arcsec southwest from the nucleus, where the simulated values are clearly higher than the observed ones. Since the S/N ratios in this region are higher than 100, the parameters provided by the PPXF method there are probably reliable; therefore, we believe that the observed discrepancies are due to limitations of our model. Although the observed $h_3$ map is noisier than the others, the main pattern of the map, which shows an anti-correlation with the $V_*$ map, was reproduced by the model. The residual map of $h_3$ does not show relevant patterns or structures.  

The parameters of the best obtained model, with $\chi_T^2 = 1.88$, are shown in Table~\ref{tbl1}. The uncertainties of these parameters were estimated using the same procedure described in \citet{der11}. First, we plotted an histogram for each parameter, but only taking into account parameters of the models with $\chi^2_T - \chi^2_{min} < 1$. After that, Gaussian functions were fitted to each one of the histograms and the square deviations of these Gaussians were taken as the uncertainties of the parameters. 

\begin{table}
\begin{center}
\caption{Parameters of the best model obtained to reproduce the maps of the stellar kinematic parameters of the data cube of NGC 4258.\label{tbl1}}
\begin{tabular}{cc}
\hline
Parameter      & Value         \\
\hline
$M_{\bullet}(disc)$  & $(2.8 \pm 1.0) \times 10^7 M_{\sun}$ \\
$i_{disc}$  & $42\degr \pm 4\degr$ \\
$M/L_I$  & $2.0 \pm 0.5$ \\
$\sigma_d$  & $80 \pm 2$ km s$^{-1}$ \\
$\sigma_b$  & $223 \pm 14$ km s$^{-1}$ \\
\hline
\end{tabular}
\end{center}
\end{table}

\section{Discussion and comparison with previous studies}

The analysis of the emission line spectrum and of the stellar kinematics in the NIFS data cube of the central region of NGC 4258 revealed relevant information about this object. The fact that we detected a broad component in the Br$\gamma$ emission line is consistent with the classification of Seyfert 1.9 of this galaxy, given by Ho et al. (1997a). Actually, \citet{ho97b} fitted the [N \textsc{II}] $\lambda \lambda 6548,6583$ + H$\alpha$ lines with a sum of Gaussian functions and detected a broad component of H$\alpha$, with $FWHM_{H\alpha} \sim 1700$ km s$^{-1}$, which is very similar to the value of $FWHM_{Br\gamma}(broad) = 1600 \pm 29$ km s$^{-1}$, obtained in this work for Br$\gamma$. The consistency between these values suggests that the broad component of Br$\gamma$ detected here is real and not the result of possible imprecisions of the starlight subtraction. \citet{wil95} also detected broad lines in NGC 4258, but in polarized light. However, as already pointed out by \citet{ho97b}, the broad lines observed by \citet{wil95} have only $FWHM \sim 1000$ km s$^{-1}$ and both permitted and forbidden emission lines have this width in polarized light. Therefore, the broad components detected by us and by \citet{ho97b}, probably associated with the broad-line region of the AGN, are different from the broad lines observed by \citet{wil95}.

\begin{figure}
\begin{center}
  \includegraphics[scale=0.54]{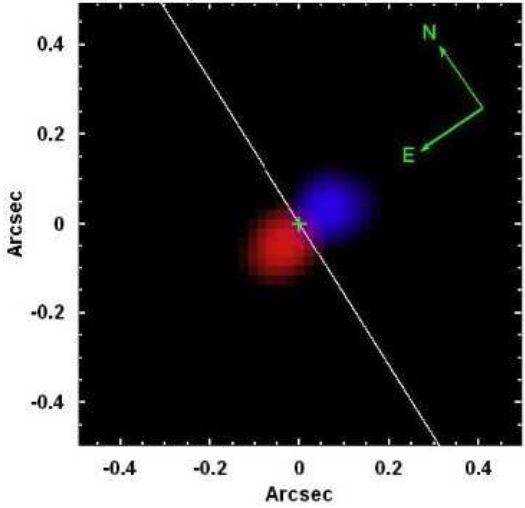}
  \caption{RB composite image, with reduced FOV, of the H$_2 \lambda 21218$ $\mu$m emission line of the data cube of NGC 4258, after the starlight subtraction. The blue and red colours represent the radial velocity ranges of -245 km s$^{-1}$ to -47 km s$^{-1}$ and 52 km s$^{-1}$ to 250 km s$^{-1}$, respectively. The white line indicates the direction of the inner radio jet in this galaxy \citep{cec00}, while the green cross corresponds to the position of the AGN.\label{fig10}}
\end{center}
\end{figure}

As mentioned in the introduction, the H$_2$O maser emitting disc in this galaxy is located between 0.16 and 0.28 pc from the SMBH \citep{miy95,her99}. The kinematics of the H$_2 \lambda 2.1218$ $\mu$m emission line detected in this work suggests the presence of a rotating molecular disc located farther from the central SMBH than the maser emitting disc. A superior limit for the diameter of this H$_2$ emitting disc is $0\arcsec\!\!.45$ (which is equivalent to 15.7 pc) and, considering the H$_2 \lambda 2.1218$ $\mu$m/Br$\gamma$ ratio, AGN excitation is the dominant mechanism in this area. The PA of the line of nodes of the $V_{H_2}$ map ($93\degr \pm 6\degr$) is compatible, at $1\sigma$ level, with that of the maser emitting disc ($PA_{maser} = 86\degr \pm 2\degr$). This is evidence that the H$_2$ emitting disc detected in this work is an extension of the inner maser emitting disc. All these rotating molecular structures are, in projection, almost perpendicular to the inner radio jet of this galaxy (with $PA_{jet} = -3\degr \pm 1\degr$). The orientation of the radio jet, relative to the molecular disc, is illustrated in the red-blue (RB) composite image, with reduced FOV, of the H$_2 \lambda 2.1218$ $\mu$m emission line shown in Fig.~\ref{fig10}. This scenario is consistent with the idea of the radio jet being collimated by a nearly edge-on inner accretion disc, which is coplanar with the rotating molecular outer structures around the SMBH. It is worth mentioning that the PA of the observed inner H$_2$ disc is not compatible, even at 3$\sigma$ level, with the PAs of the outer gas discs obtained by \citet{kru74} ($PA_{galaxy} = 146\degr$), by \citet{alb80} ($PA_{galaxy} = 150\degr$) and by \citet{saw07} ($PA_{galaxy} = 160\degr$, indicating a misalignment between the inner and outer gas discs in this galaxy.

The model of a thin stellar circular disc reproduced the main features of the $V_*$, $\sigma_*$ and $h_3$ maps. The SMBH mass resulting from this procedure ($M_{\bullet}(disc) = 2.8 \pm 1.0 \times 10^7$ $M_{\sun}$) is compatible, at $1\sigma$ level, with the most precise determination of the SMBH mass in this galaxy ($M_{\bullet}(maser) = 3.78 \pm 0.01 \times 10^7$ $M_{\sun}$; Herrnstein et al. 2005), obtained from the measurements of the rotation curve traced by the H$_2$O maser emission, assuming an inclination-warped disc. This certainly indicates that, although somewhat simplistic, our model can provide a reliable estimate for the mass of the SMBH in a galaxy. Our value of $M_{\bullet}(disc)$ is also compatible, at $1\sigma$ level, with the masses of the SMBH in NGC 4258 determined by \citet{sio09} ($M_{\bullet} = 3.3 \pm 0.2 \times 10^7$ $M_{\sun}$) and by \citet{dre15} ($M_{\bullet} = 4.8^{+0.8}_{-0.9} \times 10^7$ $M_{\sun}$). 

The PA of the line of nodes corresponding to the stellar disc in the data cube ($PA_{V_*} = 145\degr \pm 5\degr$) is compatible, at 1$\sigma$ level, with the PA of the line of nodes of the outer ionized gas disc ($PA_{galaxy} = 146\degr$) and of the outer H \textsc{I} disc ($PA_{galaxy} = 150\degr$), observed by \citet{kru74} and \citet{alb80}, respectively. The $PA_{V_*}$ value is also compatible, at 3$\sigma$ level, with the PA of the line of nodes of the outer CO disc ($PA_{galaxy} = 160\degr$), observed by  \citet{saw07}. On the other hand, the inclination of the stellar disc obtained by us ($i_{disc} = 42\degr \pm 4\degr$) is not compatible, even at $3\sigma$ level, with the inclination of the outer gas discs determined by \citet{kru74} ($i_{galaxy} = 72\degr$), by \citet{alb80} ($i_{galaxy} = 72\degr$) and by \citet{saw07} ($i_{galaxy} = 65.6\degr$). This certainly suggests a misalignment between the inner stellar disc and the outer gas dics. The $PA_{V_*}$ value is not compatible, even at 3$\sigma$ level, with the PA of the molecular H$_2$ disc observed in this work ($PA_{H_2} = 91\degr \pm 5\degr$), which indicates that the inner molecular H$_2$ disc (and, consequently, the maser emitting disc, which is co-aligned with the H$_2$ disc) and the stellar disc are also misaligned.   

A kinematic decoupling between the stellar and gas components has been observed in many galaxies \citep{sar06,dav11,kat14}. Due to angular momentum conservation, gas resulting from internal processes, such as stellar mass loss, would probably be kinematically aligned with the stars \citep{dav11}. Therefore, the misalignment between the kinematics of the gas and of the stars is usually interpreted as being due to external processes (e.g. mergers, cold accretion from the intergalactic medium). The gas provided by external processes should, with time, relax into one of the rotation axes of the galaxy and become co- or counter-rotating with the stars. However, \citet{van15} determined, in a theoretical study, that the gas disc can remain misaligned with the stellar disc for a few Gyr. As a consequence, the detection of misaligned stellar and gas discs in galaxies is not unexpected.

As already discussed in \citet{men15b}, there are more detailed dynamical models, such as the Schwarzschild method \citep{sch79} and the Jeans method (described by Cappellari 2008), that can be used to reproduce the stellar kinematic properties and to estimate the SMBH mass. In the case of the Schwarzschild method, the gravitational potential is determined, based on the brightness profile of the galaxy, and then a superposition of orbits (from an orbit library) is performed, in order to reproduce the observed kinematics. On the other hand, the Jeans method uses the solution of the Jeans equations to determine the gravitational potential and to reproduce the kinematics around the SMBH. The former method usually involves the assumption that the galaxy is axisymmetric or triaxial, while the latter normally assumes axisymmetry. A good way to compare the model of a thin rotating disc with more general and detailed approaches, such as the Schwarzschild and Jeans methods, is to discuss, in detail, the pros and cons of each one. One of the main pros of the Jeans method is its easy way to deal with the Jeans equations (especially when axisymmetry is assumed). In addition, this approach requires only the knowledge of the first two moments of the velocity distribution. On the other hand, one of the cons of the Jeans method is that it can result in a negative distribution function of the stars, which, of course, does not correspond to a physical situation. Besides that, the restrictive hypothesis of axisymmetry is also a con, as this assumption is not valid for asymmetric nuclei. A good example is the nucleus of M31 (the Andromeda galaxy), which has an eccentric rotating stellar disc around the SMBH. Such eccentric disc results in a double nucleus \citep{tre95,pei03,sal04}. The Jeans method, with the assumption of axisymmetry, cannot be applied to this object. In \citet{men13}, we discovered an eccentric H$\alpha$ emitting disc in the nuclear region of M31. With our thin-disc model, we reproduced the main aspects of its kinematics and obtained a SMBH mass ($5.0^{+0.8}_{-1.0} \times 10^7$ $M_{\sun}$) compatible, at 1$\sigma$ level, with the most precise estimates resulting from the modelling of the eccentric inner stellar disc in this galaxy \citep{sal04}. In \citet{men15b}, we also detected an eccentric stellar disc around the nucleus of M104 (the Sombrero galaxy). Therefore, similar to M31, we could not apply the axisymmetric Jeans method to the nuclear region of M104, but the estimate for the SMBH mass we obtained with our thin-disc model ($9.0 \pm 2.0 \times 10^8$ $M_{\sun}$) was compatible, at 1$\sigma$ or 2$\sigma$ levels, with the estimates obtained by many previous studies \citep{kor88,ems94,kor96,jar11,mag98}. It is worth mentioning that our thin-disc model also provided reliable estimates for the eccentricities of the rotating discs in the nuclear regions of M31 and M104.

The pros of the Schwarzschild method are as follows: it is applicable to a greater variety of possible geometries (such as triaxiality, although most studies still consider axisymmetry), it ensures a positive distribution function and it is less subject to systematic errors. On the other hand, this method is computationally expensive and, due to its complexity, may be subject to some degeneracies and difficulties for the convergence. This may be particularly problematic for asymmetric nuclei. The Schwarzschild method has already been applied, for example, to the double nucleus of M31 \citep{sam02}; however, the stellar kinematics was not properly fitted, which reflects the difficulty of applying the Schwarzschild method to a complex area as the nuclear region of M31. It is important to emphasize that the degeneracies and problems of convergence of the Schwarzschild method do not invalidate or compromise previous results obtained from successful applications of this method, but only can make it difficult to apply the method in certain situations, such as asymmetric nuclear regions (possibly containing eccentric stellar discs).

As mentioned before, one of the pros of the thin-disc model is the fact that it does not require assumptions about the geometry, such as axisymmetry. Its simplicity also makes it less subject to degeneracies and problems of convergence. Considering that, we can say that the main advantage of this method is its applicability to complex and asymmetric systems, which can be more difficult to be modelled using more general and detailed methods. Another pro of the thin-disc model is the values of $\sigma_*$ obtained separately for the disc and for the inner part of the bulge, which can be useful in a variety of studies about the influence of the central SMBH on these values of $\sigma_*$. A con of the thin-disc model is certainly the larger uncertainty obtained for the mass of the SMBH, although the result obtained in this work, and also in previous works (Menezes et al. 2013; Menezes \& Steiner 2015), is compatible with those obtained with more detailed models, at 1$\sigma$ or 2$\sigma$ levels. Another con (probably the most important) of our model is the assumption of a thin disc, instead of a thick one, which may not be appropriate in some situations. However, even in such cases, the model of a thin disc may be sufficient to reproduce the main aspects of the maps of the kinematic parameters (such as $V_*$ and $\sigma_*$) and also to provide a reliable estimate for the SMBH mass. When the observed rotating disc is not sufficiently thin, the discrepancies relative to the model are more evident when the disc is nearly edge-on and in areas located farther from the plane of the disc. This was the case for the nuclear region of M104 \citep{men15b}. Even so, as mentioned before, the SMBH mass resulting from our model was consistent with the values obtained by previous works.

The use of a model with a thin rotating disc can be attempted every time the velocity map of the nuclear region of a galaxy reveals a pattern consistent with rotation. Possible discrepancies between the modelled and observed values (especially if the disc is approximately edge-on) may indicate that the disc is not sufficiently thin to be reproduced by this model. However, even in these cases, this model will probably be able to reproduce the main aspects of the maps of the kinematic parameters and also to provide a reliable value for the SMBH mass. On the other hand, there is no doubt that the higher precision of the results obtained with other procedures, such as the Schwarzschild method, is a very important point that must be taken into account. Considering all of that, we believe that, whenever possible, more general and detailed methods should be attempted, especially when the hypothesis of axisymmetry is adequate. However, in more complex situations or when convergence problems make it difficult to obtain a result, simpler procedures, such as the thin-disc model, should be used, as they will provide reliable results as well.

\section{Summary and conclusions}

We analysed the emission line properties and also the stellar kinematics in a data cube of the central region of NGC 4258, obtained, in the \textit{K} band, with NIFS at the Gemini-north telescope. The main conclusions of this work are the following.

\begin{itemize}

\item The nuclear spectrum, after the starlight subtraction, reveals only the H$_2 \lambda 2.1218$ $\mu$m and Br$\gamma$ emission lines. Gaussian fits applied to Br$\gamma$ show a broad component with $FWHM_{Br\gamma}(broad) = 1600 \pm 29$ km s$^{-1}$, which is consistent with the classification of this galaxy as Seyfert 1.9 (Ho et al. 1997a).

\item The Br$\gamma$ emission line does not show any apparent kinematics. However, the kinematics of H$_2 \lambda 2.1218$ $\mu$m indicates the presence of a rotating molecular disc, with a superior limit for its diameter of 15.7 pc. The consistency between the PA of the line of nodes of this possible disc ($93\degr \pm 6\degr$) and that of the inner H$_2$O maser emitting disc ($PA_{maser} = 86\degr \pm 2\degr$) suggests that the former structure is an extension of the latter.

\item The inner radio jet in NGC 4258 \citep{cec00} is, in projection, almost perpendicular to the H$_2$O maser emitting disc and to the H$_2$ emitting structure detected in this work, which is consistent with the idea of such radio jet being collimated by a nearly edge-on inner accretion disc coplanar with the rotating molecular outer structures around the SMBH.

\item The model of a thin stellar circular disc reproduced the main features of the $V_*$, $\sigma_*$ and $h_3$ maps. In addition, the SMBH mass obtained with this model ($M_{\bullet}(disc) = 2.8 \pm 1.0 \times 10^7$ $M_{\sun}$) is compatible, at $1\sigma$ level, with the estimate provided by the measurements of the H$_2$O maser emission ($M_{\bullet}(maser) = 3.78 \pm 0.01 \times 10^7$ $M_{\sun}$; Herrnstein et al. 2005), which is one of the most precise mass determinations for a SMBH in a galaxy. 

\item Based on the results provided by the dynamical modelling and also on the results obtained in Menezes et al. (2013) and Menezes \& Steiner (2015), we conclude that the relatively simple model of a thin rotating disc can provide a reliable estimate for the mass of the SMBH in a galaxy. Such procedure can be attempted when a pattern of rotation is detected around the nucleus of a galaxy and it may be especially useful to analyse complex and asymmetric nuclei, where more detailed and general methods may not be applicable.

\end{itemize}

\section*{Acknowledgements}

Based on observations obtained at the Gemini Observatory (processed using the \textsc{Gemini IRAF} package), which is operated by the Association of Universities for Research in Astronomy, Inc., under a cooperative agreement with the NSF on behalf of the Gemini partnership: the National Science Foundation (United States), the National Research Council (Canada), CONICYT (Chile), the Australian Research Council (Australia), Minist\'{e}rio da Ci\^{e}ncia, Tecnologia e Inova\c{c}\~{a}o (Brazil) and Ministerio de Ciencia, Tecnolog\'{i}a e Innovaci\'{o}n Productiva (Argentina).  This work was supported by Coordena\c{c}\~ao de Aperfei\c{c}oamento de Pessoal de N\'ivel Superior (CAPES) and Funda\c{c}\~ao de Amparo \`a Pesquisa do Estado de S\~ao Paulo (FAPESP; under grant 2011/51680-6). We thank the anonymous referee for valuable comments about this paper.

\end{document}